\documentclass[pra,10pt,twocolumn,superscriptaddress,floatfix,showpacs]{revtex4-2}

\usepackage[utf8]{inputenc}
\usepackage[T1]{fontenc}     
\usepackage[british]{babel}  
\usepackage[sc,osf]{mathpazo}
\usepackage[scaled=0.86]{berasans}  
\usepackage[colorlinks=true, citecolor=blue, urlcolor=blue]{hyperref}  
\usepackage{graphicx} 
\usepackage{subfig}
\usepackage[babel]{microtype}  
\usepackage{amsmath,amssymb,amsthm,bm,amsfonts,mathrsfs,bbm} 

\usepackage{xspace}  
\usepackage{pgfplots}
\usepackage{xcolor,colortbl}
\def\ba{\begin{equation}}
\def\ea{\end{equation}}
\def\bea{\begin{eqnarray}}
\def\eea{\end{eqnarray}}
\def\ben{\begin{equation*}}
\def\een{\end{equation*}}
\def\bean{\begin{eqnarray*}}
\def\eean{\end{eqnarray*}}
\def\bma{\begin{mathletters}}
\def\ema{\end{mathletters}}
\def\bi{\begin{itemize}}
\def\ei{\end{itemize}}

\newcommand{\be}{\begin{equation}}
\newcommand{\ee}{\end{equation}}

\newcommand{\kommentar}[1]{}

\newcommand{\forget}[1]{}

\begin{document}

\title{Persistency Of Non $n$-Local Correlations In Noisy Linear Networks}
\author{Kaushiki Mukherjee*}
\email{kaushiki.wbes@gmail.com}
\affiliation{Department of Mathematics, Government Girls' General Degree College, Ekbalpore, Kolkata-700023, India.}
\author{Indranil Chakrabarty}
\email{indranil.chakrabarty@iiit.ac.in}
\affiliation{Center for Quantum Science and Technology, International Institute of Information Technology, Gachibowli, Hyderabad 500 032, Telangana, India.\\
Center for Security Theory and Algorithmic Research, International Institute of Information Technology, Gachibowli, Hyderabad 500 032, Telangana, India.}
\author{Ganesh Mylavarapu}
\email{ganesh.mylavarapu@research.iiit.ac.in}
\affiliation{Center for Quantum Science and Technology, International Institute of Information Technology, Gachibowli, Hyderabad 500 032, Telangana , India.}

\begin{abstract}
Linear $n$-local networks are compatible with quantum repeaters based entanglement distribution protocols. Different sources of imperfections such as error in entanglement generation, communication over noisy quantum channels and imperfections in measurements result in decay of quantumness across such networks. From practical perspectives it becomes imperative to analyze non classicality of quantum network correlations in presence of different types of noise.  Present discussion provides a formal characterization of non $n$-local feature of quantum correlations in noisy network scenario. In this context, persistency of non $n$-locality has been introduced. Such a notion helps in analyzing decay of non $n$-local feature of network correlations with increasing length of the linear network in presence of one or more causes of imperfections.
\end{abstract}

\maketitle

	
\section{Introduction}\label{intro}
Formulation of the Einstein-Podolsky-Rosen (EPR) paradox\cite{Ein} points out inexplicability of quantum predictions in terms of only local hidden variable models. Such an impossibility in turn gives rise to the notion of nonlocality\cite{bel}. Quantum nonlocality serves as a resource in multifaceted practical tasks\cite{may,bar,bru,pir,aci,col,brunrev}. Over the past few years study of nonlocality has been extended beyond paradigm of standard Bell scenario. Manifestation of nonlocal network correlations has been a recent trend of analysis in the field of quantum information theory\cite{birev}.\\
\par Unlike standard Bell-CHSH scenario, any measurement scenario compatible with network topology involves multiple distant sources. Each of the sources distributes physical systems to a subset of distant observers. In case all the sources in the entire network are independent of each other($n$\textit{-local assumption}), non $n$\textit{-local} correlations may emerge under suitable measurement contexts.\cite{birev}. The simplest of this type of networks, commonly known as \textit{bilocal} network(see Fig.\ref{fig1} for $n$$=$$2$) was first introduced in \cite{BRA} followed by a vivid analysis in \cite{BRAN}. Keeping pace with utility of quantum networks in various information processing tasks\cite{indr1,indr3,internet1,internet2,lee}, study of $n$-local networks has witnessed multi directional development\cite{bilo1,km1,bilo2,km2,km3,bilo4,km4,bilo3,km5,bilo7,bilo5,ejm,bilo6,km6}.\\
\par Assumption of source independence adds new physical insights in analyzing non classical behavior of quantum network correlations. For example, consider a $(n+1)$-partite entanglement swapping network(see Fig.\ref{fig1}) involving $n$ independent sources $\mathcal{S}_i(i$$=$$1,2,..,n)$ arranged in a linear fashion. Each source distributes a two qubit entangled state between a pair of parties(detailed discussion in \ref{pre1}). All the parties are thus not receiving qubits from a single source. Hence, unlike standard Bell scenario, initially they do not share any common past. Moreover some of the parties perform a single measurement. This leads to another striking difference with standard Bell experiment where each party must randomly and independently choose from a collection of two or more inputs\cite{brunrev}$.\,$ $n$-local assumption thus reduces requirements for exploiting non classicality in quantum networks\cite{gis1,BRA,BRAN}.\\
\par Quantum repeaters form building blocks of any network meant for distributing entanglement between distant observers across a large length of quantum channel\cite{indr1}. Now entanglement swapping forms the basis of designing quantum repeater networks. So any such network structure can be considered as a $n$-local network\cite{birev}. In ideal scenario, under suitable measurement contexts, nonlocality in terms of non $n$-locality is thus generated in the network. However in practical situations various factors of difficulties such as imperfection in entanglement generation, communication over noisy quantum channels and many others hinder distribution of entanglement over the entire length of the chain. Consequently, unlike that in idealistic scenario, simulation of non $n$-local correlations in the entire network structure becomes impossible. At this junction it becomes imperative to explore for how long such non classical behavior can be observed. To facilitate the discussion we have introduced the concept of \textit{persistency} in this context. \\
\par In literature, idea of persistency has been used to characterize different types of multipartite quantum correlations\cite{per1,per2,per3}. Starting from a given $m$ partite state $\rho$(say) exhibiting some form of quantum correlation $\mathcal{C}$(say), number of parties is gradually decreased so as to find the minimum number of parties $m^{'}$(say) such that none of the possible $m^{'}$ partite reduced state exhibits $\mathcal{C}.$ $m^{'}$ is usually referred to as persistency of $\rho$ with respect to the specified quantum correlation($\mathcal{C}$). For current discussion, we have introduced concept of persistency on a different note. Here it will be used for exploiting sustainability of non $n$-local feature varying with length of a network in presence of different types of noise.
\par We have analyzed generation of non $n$-local correlations in presence of various sources of imperfection. For our purpose we have considered $n$-local linear\cite{km1} networks. There may be error in entanglement generation at the sources. Distribution of qubits may then occur over noisy channels. Also the observers may be using local imperfect measurement devices at their end. We have considered all such potential sources of errors. For rest of our work, $n$-local networks affected by at least one such type of imperfection are referred to as \textit{noisy $n$-local networks.} To characterize non $n$-local correlations in such networks we put forward the notion of \textit{persistency of non $n$-locality}.\\
\par Firstly, we have derived the non $n$-locality detection criterion for noisy networks. That criterion is further used to develop the concept of persistency. First concept of persistency has been introduced in presence of single noise factor at a time. Then the notion has been generalized for more practical situations when the network is affected by two or more noise factors simultaneously.\\
\par Rest of the work is organized as follows: Some basic preliminaries are discussed in Sec.\ref{pre}. Characterization of noisy $n$-local linear is given in Sec.\ref{noise}. Persistency of  non $n$-local correlations is studied in Sec.\ref{pers} followed by some concluding remarks in Sec.\ref{conc}.
\section{Preliminaries}\label{pre}
We first proceed to discuss some basic pre-requisites to be used in forthcoming sections.
\subsection{Density Matrix Representation Of Arbitrary Two Qubit State}
Let $\varrho$ denote an arbitrary two qubit state. Density matrix of $\varrho$ in terms of Bloch parameters is given by:
\begin{equation}\label{st4}
\small{\varrho}=\small{\frac{1}{4}(\mathbb{I}_{2}\times\mathbb{I}_2+\vec{a}.\vec{\sigma}\otimes \mathbb{I}_2+\mathbb{I}_2\otimes \vec{b}.\vec{\sigma}+\sum_{j_1,j_2=1}^{3}w_{j_1j_2}\sigma_{j_1}\otimes\sigma_{j_2})},
\end{equation}
where $\vec{\sigma}$$=$$(\sigma_1,\sigma_2,\sigma_3), $ $\sigma_{j_k}$ denote Pauli operators along three mutually perpendicular directions ($j_k$$=$$1,2,3$). $\vec{a}$$=$$(x_1,x_2,x_3)$ and $\vec{b}$$=$$(y_1,y_2,y_3)$ denote
local bloch vectors ($\vec{a},\vec{b}$$\in$$\mathbb{R}^3$) corresponding to party $\mathcal{A}$ and $\mathcal{B}$ respectively with $|\vec{a}|,|\vec{b}|$$\leq$$1$ and $(w_{i,j})_{3\times3}$ denotes correlation tensor $\mathcal{W}$(real).
Matrix elements $w_{j_1j_2}$ are given by $w_{j_1j_2}$$=$$\textmd{Tr}[\rho\,\sigma_{j_1}\otimes\sigma_{j_2}].$ \\
$\mathcal{W}$ can be diagonalized by subjection it to suitable local unitary operations\cite{luo,gam}. Simplified expression is then given by:
 \begin{equation}\label{st41}
\small{\varrho}^{'}=\small{\frac{1}{4}(\mathbb{I}_{2}\times\mathbb{I}_2+\vec{\mathfrak{a}}.\vec{\sigma}\otimes \mathbb{I}_2+\mathbb{I}_2\otimes \vec{\mathfrak{b}}.\vec{\sigma}+\sum_{j=1}^{3}t_{jj}\sigma_{j}\otimes\sigma_{j})},
\end{equation}
$T$$=$$\textmd{diag}(t_{11},t_{22},t_{33})$ denote the correlation matrix in Eq.(\ref{st41}) where $t_{11},t_{22},t_{33}$ are the eigen values of $\sqrt{\mathcal{W}^{T}\mathcal{W}},$ i.e., singular values of $\mathcal{W}.$

\subsection{$n$-local Linear Networks}\label{nlocal1}
\begin{center}
\begin{figure}
\includegraphics[width=3.9in]{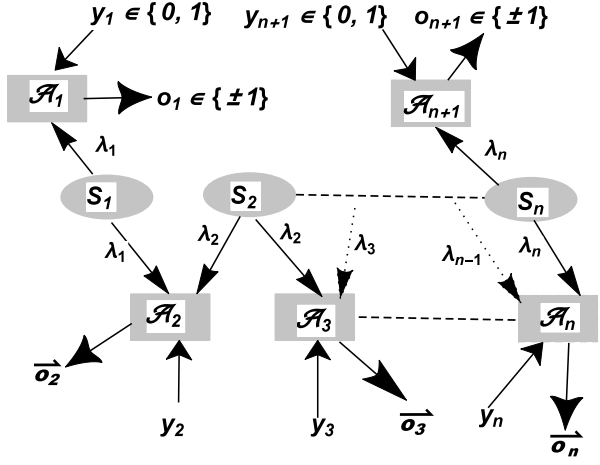} \\
 \caption{\emph{ Schematic diagram of $n$-local linear network\cite{km1}}}
\label{fig1}
 \end{figure}
 \end{center}

Consider a network with $n$ sources $\mathcal{S}_1,\mathcal{S}_2,...\mathcal{S}_n$ and $n+1$ parties $\mathcal{A}_1,\mathcal{A}_2,...,\mathcal{A}_{n+1}$ arranged in a linear pattern(see Fig.\ref{fig1}). $\forall i$$=$$1,2,...,n,$ source $\mathcal{S}_i$ independently distributes physical systems(characterized by $\lambda_i$) to $\mathcal{A}_i$ and $\mathcal{A}_{i+1}.$ For each of $i$$=$$2,3,..,n,$ $ \mathcal{A}_i$ receives two particles and is referred to as \textit{central} party. Each of other two parties $\mathcal{A}_1$ and $\mathcal{A}_{n+1}$ receives one particle and is referred to as \textit{extreme} party. $\mathcal{S}_i$ is characterized by variable $\lambda_i.$ As sources are independent, joint distribution of $\lambda_1,...,\lambda_n$ is factorizable:
\begin{equation}\label{tr1}
    \rho(\lambda_1,...\lambda_n)=\Pi_{i=1}^n\rho_i(\lambda_i)
 \end{equation}
where $\forall i,\,\rho_i$ denotes the normalized distribution of $\lambda_i.$ Eq.(\ref{tr1}) represents $n$-local constraint. \\
$\forall i$$=$$2,3,...n-1$ party $\mathcal{A}_i$ performs single measurement $y_i$ on joint state of two subsystems received from $\mathcal{S}_{i-1}$ and
$\mathcal{S}_{i}.$
Each of $\mathcal{A}_1$ and $\mathcal{A}_{n+1}$ selects from a collection of two dichotomous inputs. $n+1$
partite network correlations are local if:
\begin{eqnarray}\label{tr2}
&&\small{p(o_1,\vec{\mathfrak{o}}_2,...,\vec{\mathfrak{o}}_n,o_{n+1}|y_1,y_{n+1})}=\nonumber\\
&&\int_{\Lambda_1}\int_{\Lambda_2}...\int_{\Lambda_n}
 d\lambda_1d\lambda_2...d\lambda_n\,\rho(\lambda_1,\lambda_2,...\lambda_n) N_1 ,\,\textmd{where}\nonumber\\
&&N_1=p(o_1|y_1,\lambda_1)\Pi_{j=2}^n p(\vec{\mathfrak{o}}_j|\lambda_{j-1},\lambda_j) p(o_{n+1}|y_{n+1},\lambda_n)\nonumber\\
&&
\end{eqnarray}
Notations appearing in Eq.(\ref{tr2}) are detailed below:
\begin{itemize}
  \item $\forall j,$ $\Lambda_j$ labels the set of all possible values of local hidden variable $\lambda_j.$
  \item $y_1,y_{n+1}$$\in\{0,1\}$ denote measurements of $\mathcal{A}_1$ and $\mathcal{A}_{n+1}$ respectively.
  \item $o_1,o_{n+1}$$\in$$\{\pm 1\}$ denote outputs of $\mathcal{A}_1$ and $\mathcal{A}_{n+1}$ respectively.
  \item $\forall j,$ $\vec{\mathfrak{o}}_j$$=$$(o_{j1},o_{j2})$ labels four outputs of input $y_j$ for $o_{ji}$$\in$$\{0,1\}$

\end{itemize}
Correlations are $n$-local if those satisfy both Eqs.(\ref{tr1},\ref{tr2}). So any set of $n+1$ partite correlations that do not satisfy both
of these constraints are termed as non $n$-local.\\
\par $n$-local inequality\cite{km1} corresponding to this network scenario is:
\begin{eqnarray}\label{ineq}
   && \sqrt{|I|}+\sqrt{|J|}\leq  1,\,  \textmd{where}\nonumber\\
 && I=\frac{1}{4}\sum_{y_1,y_{n+1}}\langle O_{1,y_1}O_2^0.....O_n^0O_{n+1,y_{n+1}}\rangle\nonumber\\
 &&   J= \frac{1}{4}\sum_{y_1,y_{n+1}}(-1)^{y_1+y_{n+1}}\langle \small{O_{1,y_1}O_2^1...O_n^1O_{n+1,y_{n+1}}}\rangle\,\,\textmd{with} \nonumber\\
&&   \langle O_{1,y_1}O_2^i.....O_n^iO_{n+1,y_{n+1}}\rangle = \sum_{\mathcal{D}}(-1)^{\mathfrak{o}_1+\mathfrak{o}_{n+1}+\mathfrak{o}_{2i}+...\mathfrak{o}_{ni}}N_2,\nonumber\\
&& \textmd{\small{where}}\,N_2=\small{p(\mathfrak{o}_1,\vec{\mathfrak{o}}_2,...,\vec{\mathfrak{o}}_n,\mathfrak{o}_{n+1}|y_1,y_{n+1})},\, i=0,1\nonumber\\
&&\textmd{\small{and}}\, \mathcal{D}=\{\mathfrak{o}_1,\mathfrak{o}_{21},\mathfrak{o}_{22},...,\mathfrak{o}_{n1},\mathfrak{o}_{n2},\mathfrak{o}_{n+1}\}
\end{eqnarray}
Violation of Eq.(\ref{ineq}) ensures non $n$-local nature of corresponding correlations.

\subsection{Quantum Linear $n$-local Network Scenario}\label{pre1}
In the $n$-local network, let $\mathcal{S}_i(i$$=$$1,2,...,n)$ generate an arbitrary two qubit state $\varrho_i.$ Each of the central parties $\mathcal{A}_i(i$$=$$2,3,...,n)$ thus receives two qubits: one of $\varrho_{i-1}$ and another of $\varrho_i.$ $\mathcal{A}_1$ and $\mathcal{A}_{n+1}$ receive single qubit of $\varrho_1$ and $\varrho_n$ respectively. Let each of the central parties perform projection in Bell basis $\{|\psi^{\pm}\rangle,|\phi^{\pm}\rangle\}$, often referred to as Bell state measurement(BSM\cite{BRAN}). Let $M_i$ denote BSM of central party $\mathcal{A}_i.$ Let each of the extreme parties perform projective measurements in any one of two arbitrary directions: $\{\vec{m}_0.\vec{\sigma},\vec{m}_1.\vec{\sigma}\}$ for $\mathcal{A}_1$ and $\{\vec{n}_0.\vec{\sigma},\vec{n}_1.\vec{\sigma}\}$ for $\mathcal{A}_{n+1}$ with $\vec{m}_0,\vec{m}_1,\vec{n}_0,\vec{n}_1$$\in$$\mathcal{R}^3.$ Under this measurement settings, non $n$-local correlations are detected by violation of Eq.(\ref{ineq}) if\cite{bilo5}:
\begin{equation}\label{tribd}
    \sqrt{\Pi_{i=1}^nt_{i11}+\Pi_{i=1}^nt_{i22}}>1
\end{equation}
with $t_{i11},t_{i22}$ denoting largest two singular values of correlation tensor ($T_i$) of $\varrho_i\,(i$$=$$1,2,...,n).$ In case Eq.(\ref{tribd})
is violated nothing can be concluded about $n$-local behavior of the correlations.
\section{Noisy $n$-local linear network}\label{noise}
Consider $n$-local linear network(Fig.\ref{fig1}). Entire procedure in the network can be divided into two phases: \textit{Preparation Phase}  and \textit{Measurement Phase}. Former phase further comprises of two parts: \textit{generation  and distribution of entanglement.} For analysis of non $n$-locality in the noisy network, errors are considered in all these stages. In case the network is used for distribution of entanglement, ideally pure entanglement is to be distributed from each source $S_i(i$$=$$1,2,...,n)$. However, errors in the preparation phase lead to distribution of mixed two qubit state $\varrho_i$ among the parties $A_{i}$ and $A_{i+1}\forall i.$\\
We first analyze the correlations considering error in measurement stage. Precisely speaking, under measurement imperfections, closed form of upper bound of $n$-local inequality(Eq.(\ref{ineq})) is derived for arbitrary two qubit states. This form is further utilized in exploiting non $n$-locality under effect of errors in preparation phase. \\
\subsection{Imperfection In Measurements}
As discussed in subsec.\ref{pre1}, each of $\mathcal{A}_2,\mathcal{A}_3,...,\mathcal{A}_n$ performs BSM. Now let the devices fail to detect particles with some probability. Let $\beta_i$$\in$$[0,1]$ characterize imperfection in measurement operator $M_i$ in the sense that it fails to detect with probability $1-\beta_i.$ Measurement operator $M_i^{noisy}$(say) of $\mathcal{A}_i$ thus turns out to be a POVM  with the elements $\{M_{ij_1j_2}^{noisy}\}$ given by:
\begin{eqnarray}\label{me1}
  M_{i,00}^{noisy}&=& \beta_i |\phi^+\rangle\langle \phi^+|+\frac{1-\beta_i}{4}\mathbb{I}_{2\times2} \nonumber\\
  M_{i,01}^{noisy} &=& \beta_i |\phi^-\rangle\langle \phi^-|+\frac{1-\beta_i}{4}\mathbb{I}_{2\times 2}\nonumber \\
  M_{i,10}^{noisy} &=& \beta_i |\psi^+\rangle\langle \psi^+|+\frac{1-\beta_i}{4}\mathbb{I}_{2\times 2}\nonumber \\
  M_{i,11}^{noisy} &=& \beta_i |\psi^-\rangle\langle \psi^-|+\frac{1-\beta_i}{4}\mathbb{I}_{2\times 2},\,\forall i=2,3,...,n
\end{eqnarray}
Now, it may be noted that in case $\mathcal{A}_i$ performs perfect BSM then $\{|\phi^{\pm}\rangle\langle \phi^{\pm}|,|\psi{\pm}\rangle\langle \psi^{\pm}|\}$ is the set of possible projectors. $\forall i$$=$$2,3,...,n,$ denoting $M_{i,00}^{ideal},M_{i,01}^{ideal},M_{i,10}^{ideal},M_{i,11}^{ideal}$ as the measurement operators corresponding to the BSM projectors, POVM elements of imperfect BSM(Eq.(\ref{me1})) can be represented as:
\begin{equation}\label{me11}
     M_{i,j_1j_2}^{noisy}= \beta_i M_{i,j_1j_2}^{ideal}+\frac{1-\beta_i}{4}\mathbb{I}_{2\times2}
\end{equation}
Similar to imperfection in measurement settings of central parties, let each of the two extreme parties also use imperfect detecting devices. For party $\mathcal{A}_1,$ let $\mu$$\in$$[0,1]$ parametrizes faulty measurement device. For single qubit projection, such a device fails to detect any output with probability $1-\mu.$ POVM resulting due to imperfection in $\vec{m}_k.\vec{\sigma}$ thus has two elements $\{P_{kj}^{noisy}\}_{j=0,1}$ given by:
\begin{eqnarray}\label{me2}
  P_{k0}^{noisy} &=& \mu \mathcal{O}^++\frac{1-\mu}{2}\mathbb{I}_{2} \nonumber \\
  P_{k1}^{noisy} &=& \mu \mathcal{O}^-+\frac{1-\mu}{2}\mathbb{I}_{2},\,k=0,1
\end{eqnarray}
where $\mathcal{O}^+(\mathcal{O}^-)$ denote projection operator corresponding to $+1(-1)$ eigen value. $\mathcal{O}^{\pm}$ denote projectors corresponding to perfect projective measurement. Labeling $P_{k0}^{ideal},P_{k1}^{ideal}$ as the projectors corresponding perfect measurement $\vec{m}_k.\vec{\sigma},$ alternate representation of POVM elements(Eq.(\ref{me2})) is given by:
\begin{equation}\label{me21}
P_{ki}^{noisy}=\mu P_{ki}^{ideal}+\frac{1-\mu}{2}\mathbb{I}_{2},\,i,k=0,1
\end{equation}
Similarly for $\mathcal{A}_{n+1},$ let $1-\nu$ denote failure probability in $\vec{n}_k.\vec{\sigma}.$ Elements of corresponding POVM are given by:
\begin{eqnarray}\label{me3}
  Q_{k0}^{noisy} &=& \nu \mathcal{Q}^++\frac{1-\nu}{2}\mathbb{I}_{2} \nonumber \\
  Q_{k1}^{noisy} &=& \nu \mathcal{Q}^-+\frac{1-\nu}{2}\mathbb{I}_{2},\,k=0,1
\end{eqnarray}
with $\mathcal{Q}^+(\mathcal{Q}^-)$ denote projection operator corresponding to $+1(-1)$ eigen value. \\
We now put forward the criterion that suffices to detect non $n$-locality when all the parties are performing imperfect measurements. For $\mathcal{A}_{n+1},$ analogue of representation given by Eq.(\ref{me21}) is:
\begin{equation}\label{me31}
Q_{ki}^{noisy}=\nu Q_{ki}^{ideal}+\frac{1-\nu}{2}\mathbb{I}_{2},\,i,k=0,1
\end{equation}

\textbf{Theorem.1:} \textit{With each source $\mathcal{S}_i$ generating an arbitrary two qubit state and all the parties performing imperfect measurements, sufficient criterion for detecting non $n$-locality in a linear $n$-local network is given by:}
\begin{equation}\label{form1}
 \sqrt{\Pi_{i=1}^nt_{i11}+\Pi_{i=1}^nt_{i22}}>\frac{1}{(\mu\nu\Pi_{j=2}^n\beta_j)^{\frac{1}{2}}}.
\end{equation}
\textit{Proof:} See Appendix.\\
Eq.(\ref{form1}) being a sufficient detection criterion, violation of the same gives no definite conclusion regarding simulation of non $n$-local correlations in corresponding noisy network. Above criterion points out the effect of the imperfection parameters over the usual non $n$-locality criterion(Eq.(\ref{tribd})). Comparing right hand side of both Eqs.(\ref{tribd},\ref{form1}) it is observed that if at least one of the detectors turns out to be imperfect with some non zero probability then that reduces chances for generation of non $n$-locality in the noisy network compared to the ideal situation. Moreover, if any of the detectors used in the network always fails to detect, i.e., corresponding success probability turns out be $0$ then above criterion(Eq.(\ref{form1})) can never be satisfied.
\subsection{Noisy Entanglement Generation}
Let us first discuss an ideal entanglement generation procedure\cite{indr1}. Without loss of any generality, we consider the ideal generation of $|\phi^-\rangle\langle \phi^-|.$ Let $\varrho$$=$$|01\rangle\langle 01|$ be the state at each source $\mathcal{S}_i.$ To generate entanglement, Hadamard gate ($\mathcal{H}$) is applied on first qubit. Considering first qubit as control qubit, C$\small{-}$NOT ($\mathcal{C}\small{-}\mathcal{NOT}$) gate is then applied resulting in generation of the Bell state $|\phi^-\rangle\langle \phi^-|$\cite{indr1}. Ideally, each of $\mathcal{S}_1,\mathcal{S}_2,..,\mathcal{S}_n$ is supposed to generate $|\phi^-\rangle\langle \phi^-|.$\\
\par However in practical situations  imperfections in preparation devices lead to generation of mixed entangled states. Such imperfection results from erroneous applications of Hadamard and/or C$\small{-}$NOT ($\mathcal{C}\small{-}\mathcal{NOT}$). At each source $\mathcal{S}_i$, let $\alpha_i$ and $\delta_i$ denote the imperfection parameters characterizing $\mathcal{H}$ and $\mathcal{C}\small{-}\mathcal{NOT}$ respectively. $\forall i$$=$$1,2,...,n,$ starting from $\varrho_i$$=$$|01\rangle\langle 01|,$ noisy Hadamard gate generates\cite{indr5}:
\begin{eqnarray}\label{had1}
\varrho^{'}_i&=&\alpha_i( \mathcal{H}\otimes\mathbb{I}_2 \varrho \mathcal{H}^{\dagger}\otimes\mathbb{I}_2)+\frac{1-\alpha_i}{2}\mathbb{I}_2\otimes\varrho_{2i},\,\textmd{with}\,\alpha_i\in[0,1]\nonumber\\
&&\textmd{and}\,\, \varrho_{2i}=\textmd{Tr}_1(\varrho_i).\nonumber\\
&=&\frac{1}{2}(|00\rangle\langle 00|+|10\rangle\langle 10|)-\frac{\alpha_i}{2}(|00\rangle\langle 10|+|10\rangle\langle 00|)\nonumber\\
&&
\end{eqnarray}
Subjection of $\varrho^{'}_i$  to noisy $\mathcal{C}\small{-}\mathcal{NOT}$ gives\cite{indr5}:
\begin{eqnarray}\label{had2}
 \varrho^{''}_i&=&\delta_i( \mathcal{C}\small{-}\mathcal{NOT}\varrho^{'}_i(\mathcal{C}\small{-}\mathcal{NOT})^{\dagger})+\frac{1-\delta_i}{4}\mathbb{I}_2\otimes \mathbb{I}_2\nonumber\\
 &=&\frac{1}{4}(\sum_{i,j=0}^1(1+(-1)^{i+j}\delta_i)|ij\rangle\langle ij|-2\alpha_i\delta_i(|11\rangle\langle 00|+\nonumber\\
&&|00\rangle\langle 11|))
\end{eqnarray}
Correlation tensor of $\varrho^{''}_i$ is $\textmd{diag}(-\alpha_i\delta_i,\alpha_i\delta_i,\delta_i).$ In case $\mathcal{S}_i$ distributes $\varrho_i^{''}$ and the parties perform imperfect measurements, non $n$-locality is observed if:
\begin{equation}\label{had3}
    \sqrt{\Pi_{i=2}^n \delta_i\beta_i\mu\nu\delta_1(1+\Pi_{j=1}^n\alpha_j)}>1\,
    \end{equation}

\subsection{Noisy Quantum Communication}
Let us now consider that communication of $\varrho_i^{''}$ from $\mathcal{S}_i$ to respective parties is occurring through noisy channels. Such a communication affects generation of non $n$-local correlations for obvious reasons. To analyze effect of such noise parameters over $n$-locality detection, we are considering a few standard noisy channels\cite{nie}. \\
\subsubsection{Amplitude Damping Channel}
$\forall i$$=$$1,2,...,n,$ let $\gamma^{amp}_i,\xi_i^{amp}$ characterize channels connecting $\mathcal{S}_i$ with $\mathcal{A}_i$ and $\mathcal{A}_{i+1}$ respectively. Two qubits of $\varrho^{''}_i$(Eq.(\ref{had2})) are thus passed through two different amplitude damping channels. Let $\varrho_i^{'''}$ denote corresponding noisy state. Any amplitude damping channel(parametrized by $\gamma^{amp},$ say) is represented by Krauss operators $|0\rangle\langle 0|$$+$$\sqrt{1-\gamma^{amp}}|1\rangle\langle 1|$ and $\sqrt{\gamma^{amp}}|0\rangle\langle 1|.$ Correlation tensor of $\varrho_i^{'''}$ is given by $\textmd{diag}(-\alpha_i\delta_i\sqrt{D_i^{amp}},\alpha_i\delta_i\sqrt{D^{amp}_i},\delta_iD^{amp}_i+\gamma^{amp}_i\xi^{amp}_i)$ where
 $D^{amp}_i$$=$$(1-\gamma^{amp}_i)(1-\xi^{amp}_i).$ Using the closed form of $n$-local bound(Eq.(\ref{form1})) under imperfect measurement context non $n$-locality is detected if:
\begin{eqnarray}\label{amp1}
  \sqrt{\Pi_{i=2}^{n} \beta_i\mu\nu\textmd{Max}(2F_1,F_2)} >1\,\,&&,\,\textmd{where}\nonumber \\
   F_1= \Pi_{j=1}^n\alpha_j\delta_j\sqrt{(1-\gamma^{amp}_j)(1-\xi^{amp}_j)} && \textmd{and}\nonumber \\
   F_2=\Pi_{j=1}^n\alpha_j\delta_j\sqrt{(1-\gamma^{amp}_j)(1-\xi^{amp}_j)}+&&\nonumber\\
   \Pi_{j=1}^n(\delta_j(1-\gamma^{amp}_j)(1-\xi^{amp}_j)+\gamma^{amp}_j\xi^{amp}_j)&&
\end{eqnarray}

\subsubsection{Phase Damping Channel}
$\forall i$$=$$1,2,...,n,$ Let $\gamma^{ph}_i,\xi_i^{ph}$ characterize channels connecting $\mathcal{S}_i$ with $\mathcal{A}_i$ and $\mathcal{A}_{i+1}$ respectively. Let $\varrho_i^{'''}$ denote corresponding noisy state. Krauss operators corresponding to phase damping channel(having noise parameter $\gamma^{ph},$ say) are given by $|0\rangle\langle 0|$$+$$\sqrt{1-\gamma^{ph}}|1\rangle\langle 1|$ and $\sqrt{\gamma^{ph}}|0\rangle\langle 1|.$ Correlation tensor of $\varrho_i^{'''}$ is given by $\textmd{diag}(-\alpha_i\delta_i\sqrt{D_i^{ph}},\alpha_i\delta_i\sqrt{D_i^{ph}},\delta_i)$ where $D^{ph}_i$$=$$(1-\gamma^{ph}_i)(1-\xi^{ph}_i).$ In this case non $n$-locality is detected if:
\begin{eqnarray}\label{phase1}
\sqrt{\Pi_{i=2}^n \beta_i\mu\nu\textmd{Max}(G_1,G_2)} >1\,\textmd{where}\qquad\qquad&&,\nonumber \\
  G_1= 2\Pi_{j=1}^n\alpha_j\delta_j\sqrt{(1-\gamma^{ph}_j)(1-\xi^{ph}_j)}\, \textmd{and}&&\nonumber \\
   G_2=\Pi_{j=1}^n\alpha_j\delta_j\sqrt{(1-\gamma^{ph}_j)(1-\xi^{ph}_j)}+\Pi_{j=1}^n\delta_j&&
  \end{eqnarray}
After analyzing non $n$-locality detection in presence of noise we next introduce the notion of persistency in this context
\section{Persistency of non $n$-locality}\label{pers}
Discussion in sec.\ref{noise} clearly points out dependence of upper bound of $n$-local inequality over noise parameters. Closer observation of different relations derived therein gives rise to the intuition that increasing length of network hinders simulation of non $n$-local correlations. Formal characterization of such an interpretation will be provided in this section. In this context, let us now consider that for each of the three categories of errors discussed in sec.\ref{noise}, noise parameters remain identical. To be precise:
\begin{itemize}
  \item $\textmd{each of $n$ noisy sources is identical:}$$(\alpha_i,\delta_i)$$=$$(\alpha,\delta)\textmd(say)$\\
  $\forall i$$=$$1,2,...,n$
  \item parties are interconnected via identical noisy quantum channels
  \item single parameter characterizing imperfection in measurements of central parties $\mathcal{A}_2,...,\mathcal{A}_n:\,\beta_2$$=$$...$$=$$\beta_n$$=$$\beta\textmd(say).$
\end{itemize}
Under such assumptions Eq.(\ref{had3}) becomes:
\begin{equation}\label{had4}
    \sqrt{\mu\nu \delta^n\beta^{n-1}(1+\alpha^{n})}>1.
\end{equation}

We now define persistency of non $n$-locality for each of three types of errors individually.
\subsection{$1^{st}$ Type of Persistency of non $n$-locality }
\textbf{Definition.1} $1^{st}$ \textit{type of persistency of non $n$-locality $\mathcal{P}_I$(say) may be defined as the maximum number($n$) of independent identical sources that can be connected so as to form a linear $n$-local network where Eq.(\ref{ineq}) detects non $n$-locality under the assumption that each source distributes two qubit mixed entangled state through noiseless quantum channels and all parties perform perfect measurements.}\\
\par Above definition can be interpreted as a measure of maximum length of an entanglement distribution network in which non $n$-local correlations can be detected when the sources fail to generate pure entanglement. The measure is given in terms of independent sources as length of any network can be specified by it. If $\mathcal{P}_I$$=$$m$ for a noisy network, addition of even a single source(generating mixed entanglement) to the network will result in generation of $(m$$+$$1)$ partite correlations whose non $(m$$+$$1)$-local feature cannot be detected by Eq.(\ref{ineq}).\\
\par
\begin{center}
\begin{figure}
\includegraphics[width=3.3in]{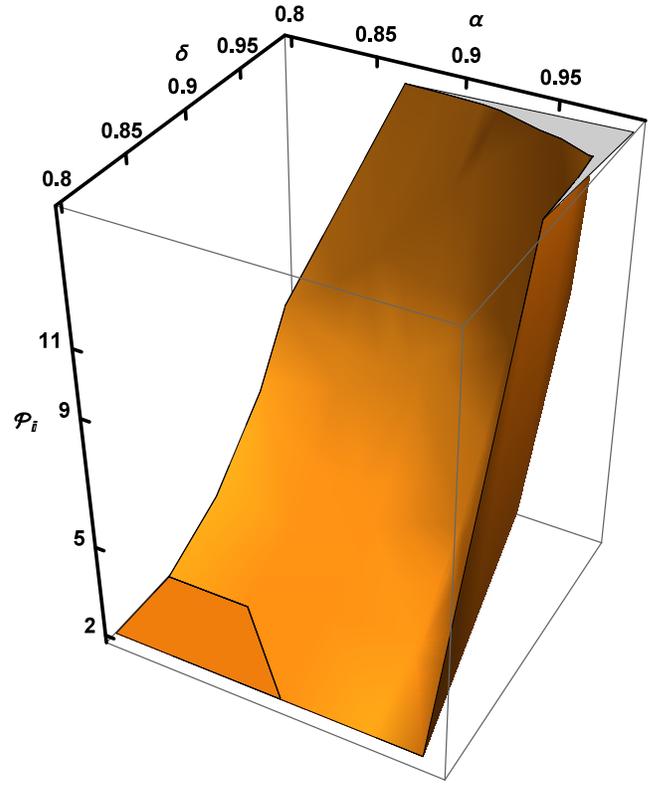} \\
 \caption{\emph{Variation of $1^{st}$ Type of Persistency of non $n$-locality with that of variables parametrizing error in entanglement generation is shown here.}}
\label{fig2}
 \end{figure}
 \end{center}

 Let us now consider a noisy network where error in entanglement generation is the only source of noise. For $\beta$$=$$\mu$$=$$\nu$$=$$1,$ Eq.(\ref{had4}) becomes:
\begin{equation}\label{pr1}
 \sqrt{\delta^n(1+\alpha^n)}>1
\end{equation}
For any specified values of $\alpha,\delta,$ $\mathcal{P}_I$ is given by:
\begin{equation}\label{pr2}
    \mathcal{P}_I=\lfloor n_I\rfloor
\end{equation}
where $n_I$ denotes the upper bound of $n$ given by Eq.(\ref{pr1}). Variation of $\mathcal{P}_I$ with that of $(\alpha,\delta)$ is plotted in Fig.\ref{fig2}. For an example, let $\alpha$$=$$\delta$$=$$0.9.$ Eq.(\ref{pr1}) gives:
\begin{equation}\label{pr3}
    n<n_I=4.567
\end{equation}
So $\mathcal{P}_I$$=$$4$ in this case.\\
\subsection{$2^{nd}$ Type of Persistency of non $n$-locality }
\textbf{Definition.2} $2^{nd}$ \textit{type of persistency of non $n$-locality $\mathcal{P}_{II}$(say) may be defined as the maximum number($n$) of independent identical sources that can be connected so as to form a linear $n$-local network where non $n$-locality is detected by Eq.(\ref{ineq}) when communication over identical noisy quantum channels is the only source of error in the network.}\\
Let $\mathcal{N}$ denote a linear $n$-local network configuration for some fixed value of $n$$=$$m\textmd{(say)}$ such that each of $m$ identical sources distributes pure entanglement through identical noisy channels and all parties perform perfect measurements. Let corresponding correlations turn out to be non $m$-local. Now let $\mathcal{N}$ be extended to an $(m$$+$$1)$-local network $\mathcal{N}^{'}$ under prevalent conditions. $\mathcal{P}_{II}$ turns out to be $m$ if Eq.(\ref{ineq}) fails to detect non $(m$$+$$1)$-locality(if any) of corresponding correlations. \\

\begin{center}
\begin{figure}
 \begin{tabular}{c}
 				\subfloat[ ]{\includegraphics[trim = 0mm 0mm 0mm 0mm,clip,scale=0.9]{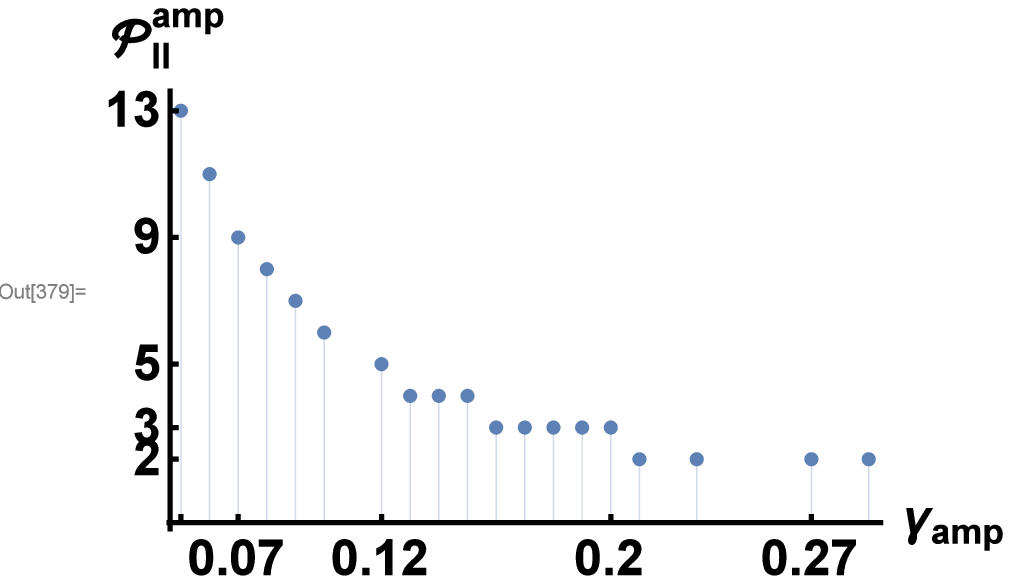}}\\
                \subfloat[]{\includegraphics[trim = 0mm 0mm 0mm 0mm,clip,scale=1.0]{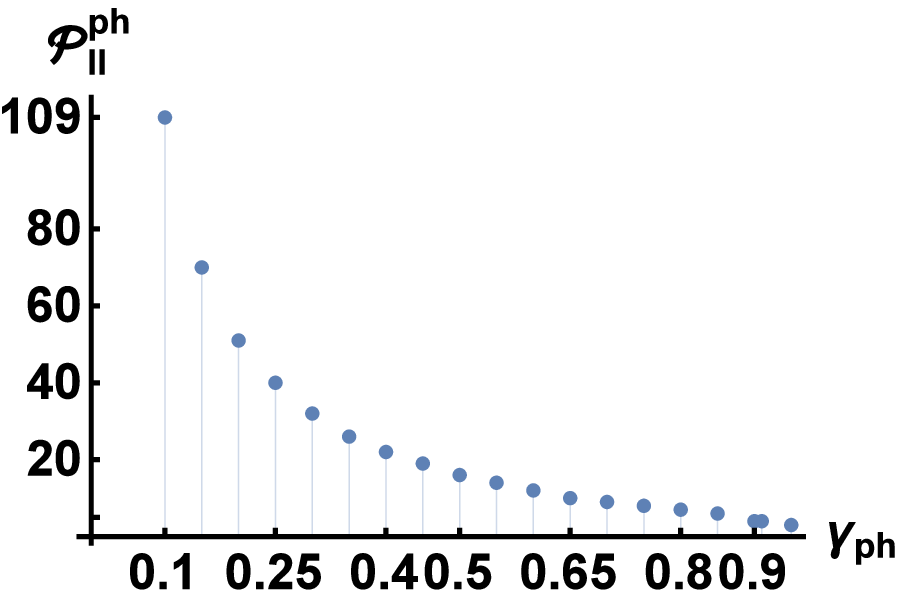}}\\
 \end{tabular}
 \caption{\emph{Decrease in $\mathcal{P}_{II}$, i.e., $\mathcal{P}_{II}^{amp}$ with increasing noise level($\gamma_{amp}$) in amplitude damping channels is plotted (sub fig.a). Similarly, for phase damping channel, $\mathcal{P}_{II}^{ph}$ vs $\gamma_{ph}$ is plotted in sub fig.b.}}
\label{fig3}
 \end{figure}
 \end{center}

Let us first consider that the sources and the parties are interconnected via identical amplitude damping channels: $\gamma^{amp}_i$$=$$\xi_i^{amp}$$=\gamma_{amp}\textmd{(say)},\forall i$$=$$1,2,,...,n.$ Non $n$-locality is detected in the network if:
\begin{eqnarray}\label{amp2}
  \sqrt{\textmd{Max}(2F_1,F_2)} &>&1,\,\textmd{where}\nonumber \\
   F_1&=& (1-\gamma_{amp})^n  \textmd{and}\nonumber \\
   F_2&=&F_1+((1-\gamma_{amp})^{2}+(\gamma_{amp})^{2})^n\nonumber\\
\end{eqnarray}
Eq.(\ref{amp2}) is obtained from Eq.(\ref{amp1}) under assumption of noisy communication as the only source of error in the $n$-local network. Eq.(\ref{amp2}) provides upper bound on the number of sources $n.$ If $n_{II}^{amp}$ denote corresponding upper bound of n, then $2^{nd}$ type of persistency of non $n$-locality is given by:$\mathcal{P}_{II}=\lfloor n_{II}^{amp}\rfloor.$ Dependence of source count($n$) and hence that of $\mathcal{P}_{II}^{amp}$ on noise parameter(see Fig.\ref{fig3}) is given by Eq.(\ref{amp2}). \\
\par Let us now consider the noisy network where communication through identical phase damping channels is the only source of noise. Setting $\gamma^{ph}_i$$=$$\xi_i^{ph}$$=\gamma_{ph}\textmd{(say)},\forall i$$=$$1,2,,...,n$ and rest of the parameters to be $1$ in Eq.(\ref{phase1}), non $n$-locality detection criterion is given by:
\begin{equation}\label{phase2}
  \sqrt{1+(1-\gamma_{ph})^n} >1.
\end{equation}
Eq.(\ref{phase2}) in turn gives $\mathcal{P}_{II}^{ph} $$=$$ \lfloor n_{II}^{ph}\rfloor$ with $n_{II}^{ph}$ denoting upper bound of $n$ given by the detection criterion(Eq.(\ref{phase2}).
Comparison of Eqs.(\ref{amp2},\ref{phase2}) indicates $\mathcal{P}_{II}^{ph}$$>$$\mathcal{P}_{II}^{amp}$ for any fixed value of noise parameter $\gamma_{amp}$$=$$\gamma_{ph}$$=$$\gamma$(see Fig.\ref{fig3}).\\

\subsection{$3^{rd}$ Type of Persistency of non $n$-locality }
\textbf{Definition.3} \textit{$3^{rd}$ type of persistency of non $n$-locality may be defined as the maximum number($n$) of independent identical sources that can be connected so that non $n$-locality is detected by Eq.(\ref{ineq}) in corresponding network under the assumption that each source distributes pure entanglement over noiseless channels and all the parties perform imperfect measurements.}\\
Let $\mathcal{P}_{III}$ denote $3^{rd}$ type of persistency of non $n$-locality. \\
\par
\begin{center}
\begin{figure}
\includegraphics[width=3.3in]{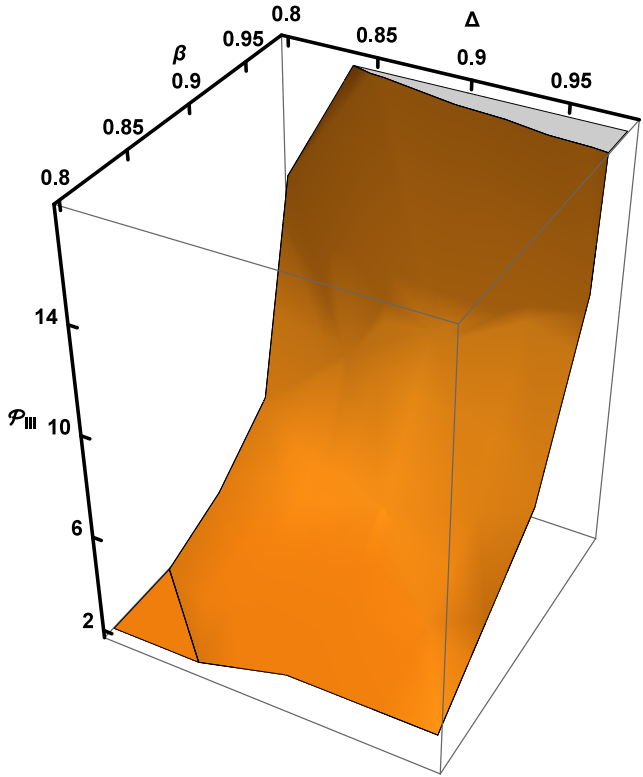} \\
 \caption{\emph{Variation of $3^{rd}$ Type of Persistency of non $n$-locality with that of imperfection in measurements is shown. Here imperfection in single qubit projective measurements of the extreme parties are parametrized by a single variable $\mu$$=$$\nu$$=$$\Delta$(say). }}
\label{fig4}
 \end{figure}
 \end{center}

 Consider $n$-local network $\mathcal{N}$ for some fixed value of $n$$=$$m\textmd{(say)}$ such that each of $m$ identical sources distributes pure entanglement through noiseless channels. Let each of the extreme parties($\mathcal{A}_1,\mathcal{A}_{m+1}$) perform imperfect projective measurements whereas each of central parties($\mathcal{A}_2,..,\mathcal{A}_m$) performs imperfect Bell basis measurement. Under such measurement contexts, let corresponding correlations turn out to be non $m$-local. Now let $\mathcal{N}$ be extended to an $(m$$+$$1)$-local network $\mathcal{N}^{'}$ by adding another identical source $\mathcal{S}_{m+1}.$ In $\mathcal{N}^{'},$ there are $m$ central parties. With all the parties performing imperfect measurements, if Eq.(\ref{ineq}) fails to detect non $(m$$+$$1)$-locality(if any) in $\mathcal{N}^{'}$ then $\mathcal{P}_{III}$$=$$m.$ \\
With imperfection in measurements considered as only source of error, non $n$-locality detection criterion Eq.(\ref{had3}) becomes:
\begin{equation}\label{had6}
    \sqrt{2\mu\nu\beta^{n-1}}>1.
\end{equation}
If $n_{III}$ denote upper bound of $n$ in Eq.(\ref{had6}) then $\mathcal{P}_{III}$$=$$\lfloor n_{III}\rfloor.$ With increase in imperfection, $\mathcal{P}_{III}$ decreases(see Fig.\ref{fig4}). For example, in case $\mu$$=$$\nu$$=$$\beta$$=$$0.9,$ non $n$-locality is detected up to $n$$=$$5.$ Hence, $P_{III}$$=$$5.$ \\
Till now persistency of non $n$-locality has been analyzed under the presence of only one type of noise at a time. However for practical purposes, it is important to study the same when at least two of the three possible factors of noise are present in the network. So generalization of the notion follows below.
\subsection{Persistency of non $n$-locality}
\textbf{Definition.4} \textit{Persistency of non $n$-locality($\mathcal{P},$say) may be defined as the maximum number($n$) of independent identical sources that can be connected to form a network such that Eq.(\ref{ineq}) detects non $n$-local correlations when at least two of the three noise factors are present in the network.}\\
\par Above definition corresponds to the most general notion of persistency. Consider $n$-local network $\mathcal{N}$ for some fixed value of $n$$=$$m\textmd{(say)}$ under the assumption that at least two of the three categories of noise are present in the network. For better understanding, without loss of any generality, let sources distribute mixed entanglement whereas parties perform imperfect measurements. Let non $m$-locality be observed in the network. On extension of $\mathcal{N}$ to an $(m$$+$$1)$-local network in presence of existing noise factors only, if Eq.(\ref{ineq}) fails to detect non $n$-locality for $n$$=$$m+1,$ then $\mathcal{P}$$=$$m$ for $\mathcal{N}.$ At this point it must be noted that in order to measure $\mathcal{P},$ extension of any network $\mathcal{N}$  to another one $\mathcal{N}^{'}$(say) by adding identical sources must be considered under the assumption that noise factors of $\mathcal{N}$ and $\mathcal{N}^{'}$ remain invariant.\\
\par Clearly $\mathcal{P}$$<$$\mathcal{P}_1,\mathcal{P}_2,\mathcal{P}_3.$ Let us now provide an example for further illustration. \\
Let us consider a noisy network where noise due to quantum communication may occur due to use of phase damp channels. Non $n$-locality detection criterion(Eq.(\ref{phase1})) is given by:
\begin{eqnarray}\label{phase5}
\sqrt{ \beta^{n-1}\mu\nu\textmd{Max}(2 G_1,G_2)} &>&1,\,\textmd{where}\nonumber \\
  G_1&=& \alpha\delta(1-\gamma_{ph})^n  \textmd{and}\nonumber \\
   G_2&=&(\alpha\delta(1-\gamma_{ph}))^n\nonumber\\
   &&+\delta^n
  \end{eqnarray}
Persistency of non $n$-locality for some specific values of noise parameters are given in Table.\ref{table:ta1}.
\begin{center}
\begin{table}[htp]
\caption{Persistency of non $n$-locality in networks under variation of noise factors. Error in communication of qubits is considered to be due to use of phase damping channels. $1^{st}$ row give persistency of non $n$-locality($\mathcal{P}$) when there is no error in entanglement generation. $2^{nd}$ row gives $\mathcal{P}$ when parties perform perfect measurements and $3^{rd}$ row gives $\mathcal{P}$ when qubits pass through noiseless channel respectively. Last row gives $\mathcal{P}$ when all three forms of errors exist in the network.}
\begin{center}
\begin{tabular}{|c|c|c|c|}
\hline
Error in&Noisy&Imperfect&$\mathcal{P}$\\
 Entanglement&Commu-&Measurements&\\
Generation&nication&&\\
\hline
$(\alpha,\delta)$$=$$(1,1)$&$\gamma$$=$$0.1$&$(\mu,\nu,\beta)$$=$$(0.94,0.93,0.92)$&$4$\\
\hline
$(\alpha,\delta)$$=$$(0.94,0.93)$&$\gamma$$=$$0.1$&$(\mu,\nu,\beta)$$=$$(1,1,1)$&$7$\\
\hline
$(\alpha,\delta)$$=$$(0.92,0.95)$&$\gamma$$=$$0$&$(\mu,\nu,\beta)$$=$$(0.92,0.94,0.95)$&$9$\\
\hline
$(\alpha,\delta)$$=$$(0.92,0.95)$&$\gamma$$=$$0.12$&$(\mu,\nu,\beta)$$=$$(0.94,0.93,0.95)$&$4$\\
\hline
\end{tabular}
\end{center}
\label{table:ta1}
\end{table}
\end{center}
\section{Discussions}\label{conc}
Characterization of non $n$-local correlations in presence of three different noise factors is provided in present work. For our purpose existing upper bound(Eq.(\ref{tribd})) of $n$-local inequality(Eq.(\ref{ineq})) has been used as the detection criterion of non $n$-locality. Persistency of non $n$-locality has been introduced for analyzing decay of non $n$-local correlations with increasing length of a noisy network. Considering persistency of $n$-locality for each of three types of noise individually, the notion has been generalized to the more practical situation when at least two of the error factors exist in the network.  \\
\par To this end, it must be pointed out that from experimental perspectives, current study turns out to be a simple form of error analysis for exploiting non $n$-locality. Though we have considered three broad categories of errors that usually occur in any entanglement swapping based network scenario, yet the discussion is oversimplified as any discussion on the technical difficulties, associated with experimental realization of quantum networks\cite{ex1}, lies beyond the scope of this manuscript. \\
\par In \cite{ex2}, the authors have pointed out multiple problems associated with physical implementation of any network configuration based on entanglement swapping. For instance, one of the most significant problem is the exponential decrease in coherence of quantum states. Such decoherence of a quantum system occurs due to long range distribution of entanglement over noisy channels and also on being subjected to operations over considerably long span of time\cite{ex1}. Methods such as entanglement purification\cite{ex3,ex4,ex5,ex6}, concentration\cite{ex3,ex6,ex6i,ex6ii,ex6iii,ex6iv}, distillation\cite{ex7,ex8,ex9} have been developed to distribute entanglement along long chain of network  With technological advancement in the field of quantum information science, practical implementations of these procedures with tolerable error rates have become possible\cite{ex5,ex6,ex6i,ex6ii,ex6iii,ex6iv,ex7,ex8,ex9}. Apart from loss in coherence, long range fiber based quantum communications becomes challenging due to photon loss, noise in photon detection and many other factors\cite{ex1}. Exponential decrease of signal-to-noise ratio with increasing length of the fiber in quantum key distribution protocols is one of the consequences of limitations over long range quantum communication. Though remarkable technological progress has been attained over years\cite{ex10,ex11,ex12,ex13,ex14,ex15}, yet, till date, there exist several limitations constraining quantum communication over large distance.\\
\par Apart from the issues mentioned above, experimentalists face several other challenges while implementing a network configuration\cite{ex1}. Hence, from experimental perspectives, it becomes important to consider at least some of these crucial factors while making any form of error analysis in a network scenario. However, our analysis has not included any such practical problem. At this junction, it is needed to be mentioned that the sole aim of present study is to introduce the notion of persistency of non $n$-locality in context of exploiting non-classicality in linear network configuration. So the study, in its present form, can be considered to be in a nascent stage which needs gradual upgradation keeping pace with present day technology. Consequently, it will be interesting to extend the study taking into consideration the practical problems. \\

\par In \cite{ex16}, entanglement swapping network has been used as Bell nonlocality activation protocol. Key role of any such protocols is to generate Bell-CHSH nonlocal quantum states starting from two or more Bell-CHSH local states. Over years such type of protocols have been generalized so as to activate different other notions of nonlocality\cite{ex17,ex18,ex19}. Now, as already discussed in sec.\ref{intro}, entanglement swapping network is a $n$-local network where each independent source distributes entangled state. Present study on persistency of non $n$-local correlations can thus be considered as one way of analyzing errors in exploiting a particular form of non-classicality(non $n$-locality) of $n+1$-partite correlations generated across any such network. Now, in place of considering correlations across the entire network configuration, it will also be interesting to study persistency of Bell nonlocality or any other notion of non-classicality of the conditional states generated at the end of any such activation network\cite{ex16,ex17}.
\par Entire analysis is limited to noisy linear $n$-local networks only. It will be interesting to exploit the same for any non linear configuration. Also in the network scenarios considered here, each source distributes two qubit entangled states. Analyzing decay of non classicality with growing imperfections in network when each of the sources generate multipartite and/or higher dimensional entangled states is a potential direction of future research.

\section{Appendix}
\textit{Proof of Theorem.1:} Let us first consider the $n$-local inequality(Eq.(\ref{ineq})) for the noisy network:
\begin{eqnarray}\label{in1}
   && \sqrt{|I_{noisy}|}+\sqrt{|J_{noisy}|}\leq  1,\,  \textmd{where}\nonumber\\
 && I_{noisy}=\frac{1}{4}\sum_{y_1,y_{n+1}}\langle O_{1,y_1}O_2^0.....O_n^0O_{n+1,y_{n+1}}\rangle_{noisy}\nonumber\\
 &&   J_{noisy}= \frac{1}{4}\sum_{y_1,y_{n+1}}(-1)^{y_1+y_{n+1}}\langle \small{O_{1,y_1}O_2^1...O_n^1O_{n+1,y_{n+1}}}\rangle_{noisy}\,\,\textmd{with} \nonumber\\
&&   \langle O_{1,y_1}O_2^i.....O_n^iO_{n+1,y_{n+1}}\rangle_{noisy} = \sum_{\mathcal{D}}(-1)^{\mathfrak{o}_1+\mathfrak{o}_{n+1}+\mathfrak{o}_{2i}+...\mathfrak{o}_{ni}}N_{noisy},\nonumber\\
&& \textmd{\small{where}}\,N_{noisy}=\small{p^{'}(\mathfrak{o}_1,\vec{\mathfrak{o}}_2,...,\vec{\mathfrak{o}}_n,\mathfrak{o}_{n+1}|y_1,y_{n+1})},\,i=0,1 \nonumber\\
&&\textmd{\small{and}}\, \mathcal{D}=\{\mathfrak{o}_1,\mathfrak{o}_{21},\mathfrak{o}_{22},...,\mathfrak{o}_{n1},\mathfrak{o}_{n2},\mathfrak{o}_{n+1}\}
\end{eqnarray}
Different symbol $p^{'}()$ has been used for probability terms so as to discriminate those arising in noisy scenario from that in ideal scenario. Let us denote the overall state in the network as $\varrho$$=$$\otimes_{l=1}^{n}\varrho_l.$
Next we consider the expectation terms given by Eq.(\ref{ineq}). Without loss of any generality let us fix $i$$=$$0$ and fix the labeling of $(y_1,y_{n+1})$ as $(0,0)$ and consider corresponding expectation term $\langle O_{1,0}O_2^0.....O_n^0O_{n+1,0}\rangle:$
\begin{widetext}
\begin{eqnarray}\label{in2}
  \langle O_{1,0}O_2^0.....O_n^0O_{n+1,0}\rangle_{noisy} &=&\sum_{\mathfrak{o}_1,\mathfrak{o}_{21},\mathfrak{o}_{22},...,\mathfrak{o}_{n1},\mathfrak{o}_{n2},\mathfrak{o}_{n+1}}(-1)^{\mathfrak{o}_1+
  \mathfrak{o}_{n+1}+\mathfrak{o}_{20}+...+\mathfrak{o}_{n0}}\small{p(\mathfrak{o}_1,\vec{\mathfrak{o}}_2,...,\vec{\mathfrak{o}}_n,\mathfrak{o}_{n+1}|
  0,0)}\nonumber\\
&=&\sum_{i,j=0}^1\sum_{g_2,h_2=0}^1\sum_{g_3,h_3=0}^1...\sum_{g_n,h_n=0}^1(-1)^{i+j+g_2+g_3+...+g_n}\textmd{Tr}[P_{0i}^{noisy}\otimes_{k=2}^n M_{k,g_kh_k}^{noisy}\,Q_{0j}^{noisy}\varrho]\, By Eqs.(\ref{me1},\ref{me2},\ref{me3})\nonumber\\
&=&\sum_{i,j=0}^1\sum_{g_3,h_3=0}^1...\sum_{g_n,h_n=0}^1 R_2\nonumber\\
where R_2&=&(-1)^{i+j+g_3+...+g_n}(\textmd{Tr}[P_{0i}^{noisy}M_{2,00}^{noisy}\otimes_{k=3}^n M_{k,g_kh_k}^{noisy}\,Q_{0j}^{noisy}\varrho]+\textmd{Tr}[P_{0i}^{noisy}M_{2,01}^{noisy}\otimes_{k=3}^n M_{k,g_kh_k}^{noisy}\,Q_{0j}^{noisy}\varrho]-\nonumber\\
  &&\textmd{Tr}[P_{0i}^{noisy}M_{2,10}^{noisy}\otimes_{k=3}^n M_{k,g_kh_k}^{noisy}\,Q_{0j}^{noisy}\varrho]-\textmd{Tr}[P_{0i}^{noisy}M_{2,11}^{noisy}\otimes_{k=3}^n M_{k,g_kh_k}^{noisy}\,Q_{0j}^{noisy}\varrho])
\end{eqnarray}
Further simplifying $R_2,$ we get:
\begin{eqnarray}\label{in3}
R_2 &=& (-1)^{i+j+g_3+...+g_n}\textmd{Tr}[P_{0i}^{noisy}(M_{2,00}^{noisy}+M_{2,01}^{noisy}-M_{2,10}^{noisy}-M_{2,11}^{noisy})\otimes_{k=3}^n M_{k,g_kh_k}^{noisy}\,Q_{0j}^{noisy}\varrho] \nonumber\\
   &=& \beta_2(-1)^{i+j+g_2+g_3+...+g_n}\textmd{Tr}[P_{0i}^{noisy}\otimes M_{2,g_2h_2}^{ideal}\otimes_{k=3}^n M_{k,g_kh_k}^{noisy}\,Q_{0j}^{noisy}\varrho] \,(using Eq.(\ref{me11}))\\
 \end{eqnarray}
Using Eq.(\ref{in3}) in Eq.(\ref{in2}), we get:
\begin{equation}\label{in4}
    \langle O_{1,0}O_2^0.....O_n^0O_{n+1,0}\rangle _{noisy} =\beta_2\sum_{i,j=0}^1\sum_{g_2,h_2=0}^1\sum_{g_3,h_3=0}^1...\sum_{g_n,h_n=0}^1(-1)^{i+j+g_2+g_3+...+g_n}\textmd{Tr}[P_{0i}^{noisy}\otimes M_{2,g_2h_2}^{ideal}\otimes_{k=3}^n M_{k,g_kh_k}^{noisy}\,Q_{0j}^{noisy}\varrho]
\end{equation}
Following similar approach of breaking sums over rest of the indices appearing in Eq.(\ref{in4}), we have
\begin{equation}\label{in5}
    \langle O_{1,0}O_2^0.....O_n^0O_{n+1,0}\rangle _{noisy} =\Pi_{i=2}^n\beta_i\mu\nu\sum_{i,j=0}^1\sum_{g_2,h_2=0}^1\sum_{g_3,h_3=0}^1...\sum_{g_n,h_n=0}^1(-1)^{i+j+g_2+g_3+...+g_n}\textmd{Tr}[P_{0i}^{ideal}\otimes_{k=2}^n M_{k,g_kh_k}^{ideal}\,Q_{0j}^{ideal}\varrho]
\end{equation}
Following same procedure for each expectation term appearing in Eq.(\ref{in1}), we get
\begin{eqnarray}\label{in6}
  && I_{noisy}=\mu\nu\Pi_{i=2}^n\beta_i I\nonumber\\
 &&   J_{noisy}=\mu\nu\Pi_{i=2}^n\beta_i J\\
\end{eqnarray}
Eq.(\ref{in1}) thus gives:
\begin{equation}\label{in7}
    \sqrt{\mu\nu\Pi_{i=2}^n\beta_i}(\sqrt{|I|}+\sqrt{|J|})=1.
\end{equation}
Eq.(\ref{in7}) is the $n$-local inequality for a linear $n$-local network where the parties perform imperfect measurements and each of the source generates an arbitrary two qubit state. As upper bound of $n$-local inequality(Eq.(\ref{ineq})) in ideal linear $n$-local network is given by Eq.(\ref{tribd}), clearly upper bound of Eq.(\ref{in7}) is given by:$\sqrt{\mu\nu\Pi_{j=2}^n\beta_j}\sqrt{\Pi_{i=1}^nt_{i11}+\Pi_{i=1}^nt_{i22}}.$ Non $n$-locality detection criterion is thus given by Eq.(\ref{form1}).\\
Proved
\end{widetext}

\end{document}